\documentclass[aps,preprint,groupedaddress,showpacs]{revtex4}
\input{epsf}

\usepackage{epsfig}
\usepackage{amsfonts}

\def\m{${\cal M}$}

\def\mf{$M_F$}

\def\ts2f{${\cal T}^{*(2F)}$}

\def\cstsa4{${\cal C}^s_{{\cal T}^{*(A_4)}}$}
\def\tsa4{${\cal T}^{*(A_4)}$}

\def\ep{$\mathbb{E}_\parallel$}
\def\es{$\mathbb{E}_\perp$}

\def\dsix{$D_6$}
\def\d*six{$D^*_6$}

\def\dfo{${{\mbox{$\bigcirc$}}\!\!\!\!\!\!\:{\mbox{3}}\,}$}
\def\ffo{${{\mbox{$\bigcirc$}}\!\!\!\!\!\!\:{\mbox{5}}\,}$}
\def\zfo{${{\mbox{$\bigcirc$}}\!\!\!\!\!\!\:{\mbox{2}}\,}$}
\begin{document}

\title{
A maximum density rule for surfaces of quasicrystals
}

\author{Z.~Papadopolos\footnote{Author
                                for correspondence:
Phone: +49 7071 29 76378; Fax: +49 7071 29 5604;
e-mail: zorka.papadopolos@uni-tuebingen.de}}
\affiliation{Institut f\"{u}r Theoretische Physik,
Universit\"{a}t T\"{u}bingen, D-72076 T\"{u}bingen,
Germany}

\author{P.~Pleasants}
\affiliation{Department of Mathematics, The University of Queensland,
St.~Lucia, QLD~4072, Australia}
\author{G.~Kasner}
\affiliation{Institut f\"{u}r Theoretische Physik,
Universit\"{a}t Magdeburg, PSF 4120, D-39016 Magdeburg,
Germany}
\author{V.~Fourn\'{e}e}
\affiliation{%Laboratoire de Science et G\'{e}nie des 
%Mat\'{e}riaux et de M\'{e}tallurgie,
Center d'Ing\'{e}nierie des
Mat\'{e}riaux, Ecole des Mines de Nancy,
F-54042 Nancy, France}
\author{T.~Cai, C.~Jenks, P.~Thiel}
\affiliation{Ames Laboratory, Iowa State University,
Ames, IA 50011, USA}
\author{J.~Ledieu, R.~McGrath}
\affiliation{Surface Science Research Centre,
The University of Liverpool, Liverpool L69 3BX, UK}

\date{\today}

\begin{abstract}
A rule due to Bravais of wide validity for crystals
is that their surfaces correspond to the densest planes
of atoms in the bulk of the material.
Comparing a theoretical model of \mbox{i-AlPdMn} with
experimental results, we find that this correspondence
breaks down and that surfaces parallel to the densest
planes in the bulk are not the most stable, i.e.~they
are not so-called bulk terminations.
The correspondence can be restored by recognizing that
there is a contribution to the surface not just from one
geometrical plane but from a layer of stacked atoms,
possibly containing more than one plane.
We find that not only does the stability of
high-symmetry surfaces match the density of the
corresponding layer-like bulk terminations but the exact
spacings between surface terraces and their degree of
pittedness 
may be determined by a simple analysis of the density 
of layers predicted by the bulk geometric model.  
\end{abstract}

\pacs{61.44.Br, 68.35.Bs, 68.37.Ef}

\maketitle

The surfaces of quasicrystals offer rich potential both
for technological applications and for fundamental
science. Their tribological properties, such as low
coefficient of friction, low surface energy and
oxidation resistance have been well documented
\cite{Dubois93} and have motivated many studies of the
clean surfaces of these materials, see
Ref.~\cite{Z02} and work cited there. 
These studies have led to significant progress 
in the past few years, especially in the case
of the aluminium-based icosahedral quasicrystals. 
The surfaces of these materials have been shown to be 
perfect ``slices'' of the bulk structure, with only 3 
distinct step heights being observed. However a 
significant  limitation of this work has been that 
there was no understanding of {\em which} bulk 
planes might be expected to be seen as surface 
terminations.  

A rule with wide validity for crystals, first suggested
by Bravais~\cite{Bravais1849} and later refined by
others, is that ``the largest facets have the densest
packing of atoms''~\cite{AW00}, usually interpreted as
meaning that, by and large, the most stable surfaces are 
those parallel to the densest atomic planes in the bulk.
It has been observed \cite{Shen00} that this does not
hold for \mbox{i-AlPdMn}, where the {\em most stable
surfaces} are orthogonal to the fivefold axes but the
{\em densest planes} in the bulk model of Boudard et 
al.~\cite{Boudard92} are orthogonal to the twofold axes.
For quasicrystals we propose modifying the
rule to use densities of thin layers of bulk planes 
instead of densities of single bulk planes.
For icosahedral quasicrystals we consider 
0.6\,{\AA} layers.
This is suggested by the fact that, whereas the
distances between neighboring high density planes in the
bulk of an ordinary crystal are 1.5-2.0\,{\AA}, in the
geometric bulk model \m~\cite{PKL,G99,MRS99,Z02} of
F~phase icosahedral quasicrystals the distances are 
0.2-1.5\,{\AA}.
We show that the observed surface structure of
\mbox{i-AlPdMn} is 
consistent with this modified rule.

\begin{figure}
\caption{ [see file 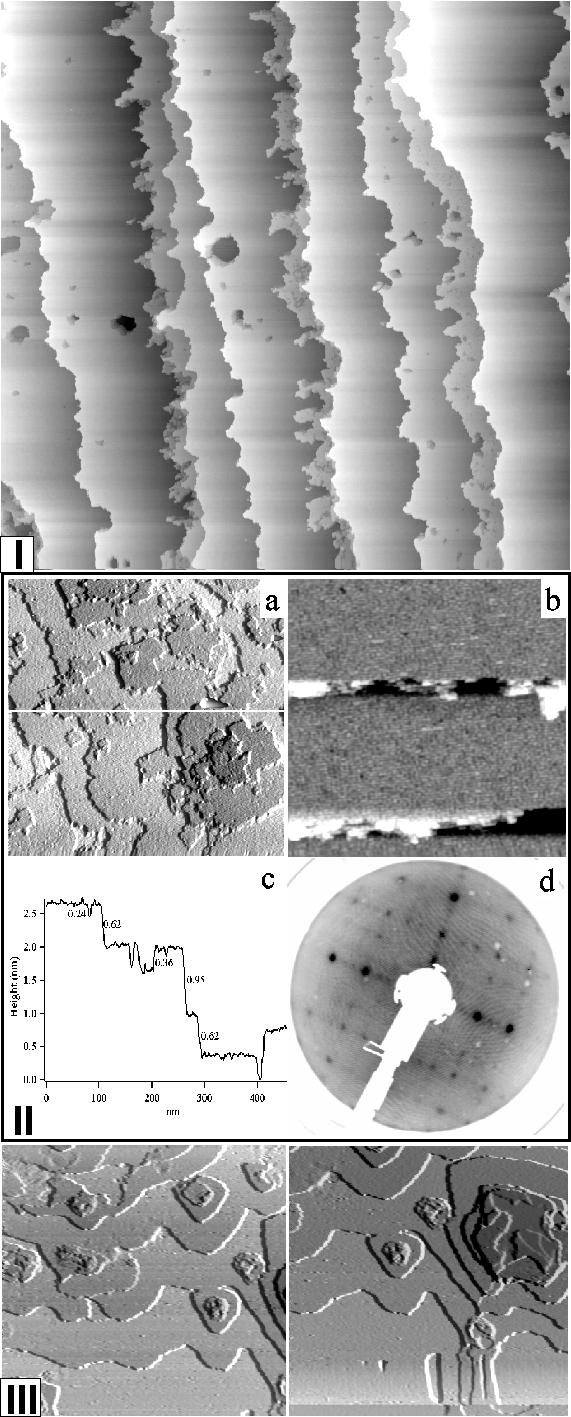]
(I) STM image ($1750\times 1750$\,nm$^2$) of the
fivefold surface of \mbox{i-AlPdMn}.
%(J Ledieu et al.),
(II) %(V~Fournee et al.)
twofold surface of \mbox{i-AlPdMn}:
(a) STM image of a terrace-stepped twofold surface
[$500\times 500$\,nm$^2$].
(b) Flattened image covering three terraces and two
steps [$160\times 160$\,nm$^2$].
(c) Height profile along the line in (a) with step
height value.
(d) LEED pattern at 50\,eV.
(III) Two STM images of a threefold surface of
\mbox{i-AlPdMn}, each $800\times 800$\,nm$^2$.
These images overlap in about half their area.
%(T Cai et al.)
}
\label{fig:fig(1)}
\end{figure}

To demonstrate this new rule we first present 
experimental evidence on the stability and appearance of 
surfaces of \mbox{i-AlPdMn} derived from scanning 
tunneling microscopy (STM) and low-energy electron 
diffraction (LEED) measurements. 
The fivefold surfaces of \mbox{i-AlPdMn}
(Fig.~\ref{fig:fig(1)}(I)) show large scale terraces and 
are stable~\cite{Shen00}.
In Ref.~\cite{G99} we showed that the intervals between
terraces on the fivefold surfaces of \mbox{i-AlPdMn} 
match a Fibonacci sequence of planes in the model \m\ 
with $S=4.08$\,{\AA} and $L=\tau S=6.60$\,{\AA}.
In Refs.~\cite{Gierer97,Z02} it was found that the
fivefold bulk terminations of \mbox{i-AlPdMn} consist
of two atomic planes 0.48\,{\AA} apart.
%
% Fig 1
%

According to Ref.~\cite{Shen00} both twofold and
threefold surfaces \textit{facet}, i.e.\ they are less 
stable.
The twofold surfaces show pitted small scale terraces,
see Fig.~\ref{fig:fig(1)}(II)(a-c).
The threefold surfaces of \mbox{i-AlPdMn} show clear
medium scale terraces with depressions (pits),
see Fig.~\ref{fig:fig(1)}(III).

\begin{figure}[]
\caption{ [see file 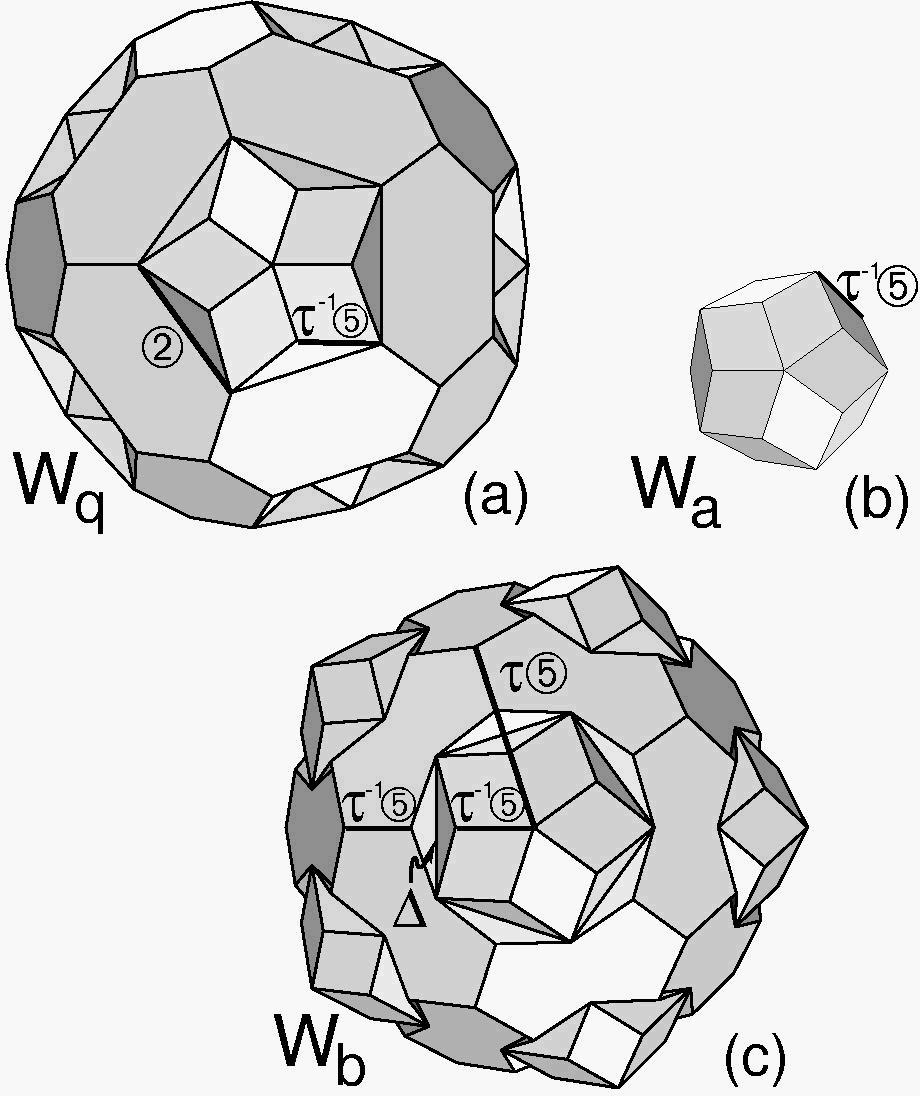]
The coding windows $W_q$, $W_a$, $W_b$
define the geometric model \m\ of atomic positions
based on the icosahedral \dsix\ module \mf:
(a) $W_q$ with edge lengths $\tau^{-1}$\ffo\ and
     \zfo\,.
(b) $W_{a}$  is a triacontahedron of edge length
     $\tau^{-1}$\ffo\,.
(c) $W_{b}$ is obtained by taking the marked tetrahedra
     away from the triacontahedron of edge length
     $\tau$\ffo\,.
}
\label{fig:fig(2)}
\end{figure}

\begin{figure}
\caption{ [see file 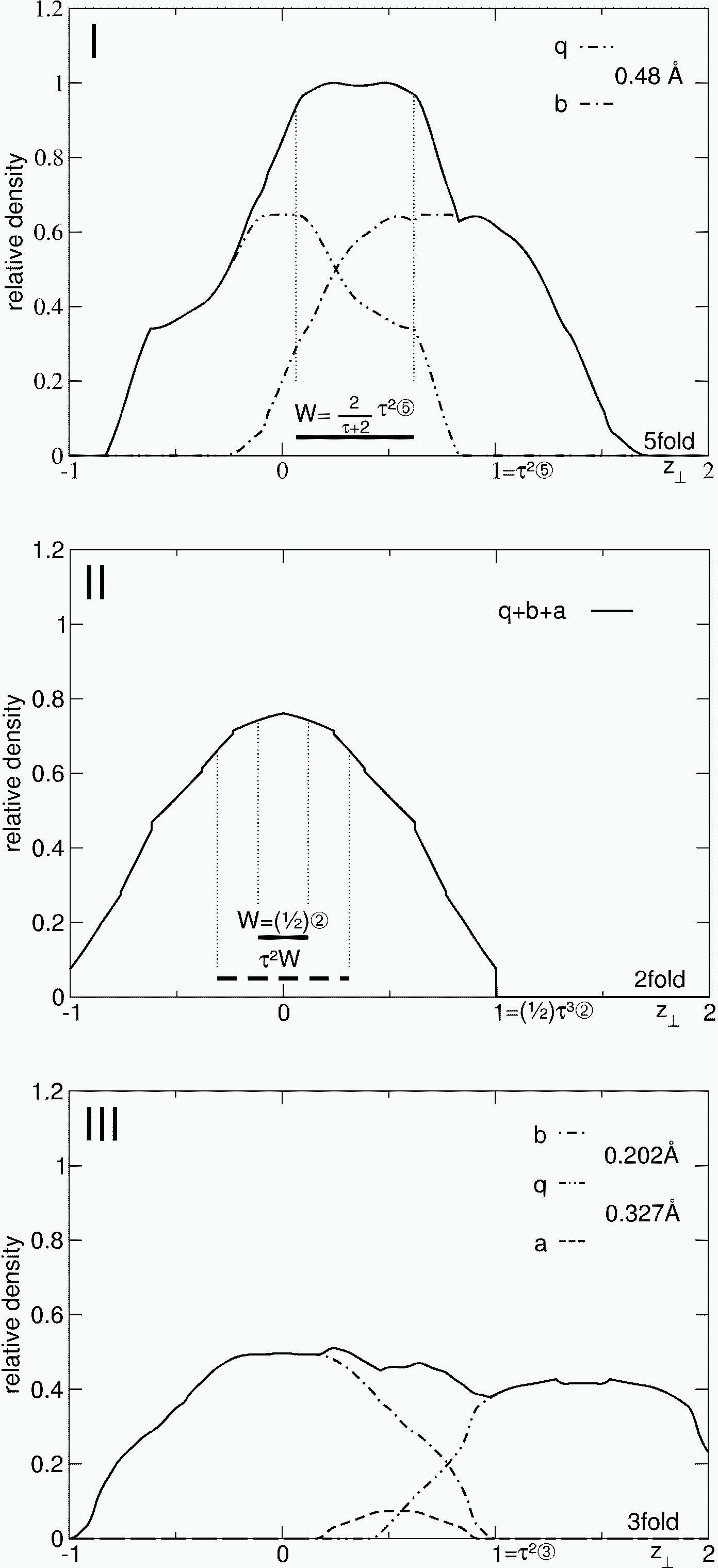]
(I) The densest fivefold layers are $(q,b)$ pairs of 
planes a distance 0.48\,{\AA} apart.
The $q$-curve is the density of atomic positions on
$q$-planes and the $b$-curve,
the density on $b$-planes.
The full curve is the sum of these densities.
The support of the plateau of the full curve defines
the terminations and is the coding window of the
Fibonacci sequence of terraces of terminations seen
on fivefold surfaces.
The height of the plateau is the density of fivefold
terminations.
(II) The densest twofold layers are single planes, which
contain atomic positions of types $q$, $b$ and $a$.
The curve graphs the density of planes along a twofold
axis in coding space.
(III) Graphs of the densities of a $(b,q,a)$ triple of
threefold planes
with spacings: $b$-plane, $0.20$\,{\AA}, $q$-plane,
$0.33$\,{\AA}, $a$-plane.
The plateau in the graph of the combined density
defines the terminations, its height giving their
density.
}
\label{fig:fig(3)}
\end{figure}

We explain this experimental evidence in terms of the
particular geometric model \m~\cite{PKL,G99,MRS99,Z02}
of the quasicrystals \mbox{i-AlPdMn}~\cite{Boudard92}.
This model is a superposition of three icosahedral
quasilattices, $q$, $a$ and $b$, of atomic positions
in the physical space \ep.
As described in Ref.~\cite{Z02} and references cited
there, there is a {\em coding space}, \es, containing
three {\em windows}, $W_q$, $W_a$ and $W_b$ (shown in
Fig.~\ref{fig:fig(2)}), and a \mbox{\em $*$-map} that
takes each point of one of the quasilattices into
a point of the corresponding window.
The \mbox{$*$-map} is not continuous:
it maps discrete unbounded quasilattices
to dense point sets bounded by a window.
Its usefulness is that it maps planes to planes and
that the density function on planes orthogonal to a
given axis, which is an erratic discrete function
($\rho(z_\parallel)$) in \ep, is a continuous
function ($\rho(z_\perp)$) in \es\ which can be
graphed as in Figs.~\ref{fig:fig(3)}(I)-(III).
For each of the five, two and threefold axes we use a 
standard distance, denoted by \ffo\,, \zfo\ and \dfo\,,
respectively.  These are related by
\dfo${}/\sqrt{3}={}$\ffo/${}\sqrt{\tau +2}={}$\zfo${}/2
=1/\sqrt{2(\tau +2)}$, where $\tau=(1+\sqrt{5})/2$.
Below we take \ffo${}=4.56$\,{\AA}---the value for
\mbox{i-AlPdMn}.
%
% Fig 2
%

For ordinary lattices the density of points in a plane
depends only on the orientation of the plane. For the
$q$, $a$, $b$ quasilattices the density $\rho $ is a
product of two factors: the {\em module factor}~\cite{PP}
that depends on the orientation of the plane
($1/\sqrt5$, $1/4$ and $1/3$, respectively, for five,
two and threefold planes)
and the {\em window factor} which is the area
of the section of the window by the corresponding
plane in coding space (see Fig.~12 in Ref.~\cite{Z02}).
We also use the fact that each plane orthogonal to a
threefold or fivefold axis contains points of one 
quasilattice only and each plane orthogonal to a twofold 
axis contains points of all three quasilattices, 
$q$, $a$ and $b$.
The top row of Table~\ref{tab:1} gives the maximum
density of planes in the main symmetry directions and
the bottom row the maximum density of terminations
(described below).
As for the similar
model with spherical windows~\cite{Boudard92} used
in Ref.~\cite{Shen00}, there are twofold planes denser 
than the densest fivefold or threefold planes even 
though experimental evidence indicates that the fivefold 
surfaces are the most stable~\cite{Shen00}.

%
% Table I
%
\begin{table}[]
\caption{
Row~1 gives the maximum atomic densities of planes
in the model \m\ and
Row~2 the maximum atomic densities of the
layer-like terminations.
}
\label{tab:1}
\begin{tabular}{c||c|c|c}
   & fivefold & twofold & threefold\\ \hline
Densest planes& 0.086\,{\AA}$^{-2}$&
0.101\,{\AA}$^{-2}$&0.066\,{\AA}$^{-2}$ \\[1mm]
Densest layers&0.133\,{\AA}$^{-2}$&
0.101\,{\AA}$^{-2}$&0.066\,{\AA}$^{-2}$
\end{tabular}
\end{table}

In the light of this, and the fact that fivefold
terminations are observed to consist of a pair of
neighboring planes, we propose a modification to the
Bravais rule to take into account close neighboring
planes. 
We consider bulk {\bf layers} of a
quasicrystal, unions of close neighboring planes 
(for icosahedral quasicrystals, modeled by \m, 
within a 0.6\,{\AA} range), 
and define a bulk {\bf termination}
to be a layer with close to maximum density among all
parallel layers.
We then expect the more stable surfaces to correspond
to the denser terminations.
To test this we now describe the terminations of our
model \m\ in the five, two and threefold directions 
and compare them with the experimental data.
In each of these directions three different inter-plane
distances occur, in each case in the ratios
$1:\tau:\tau^2$.
In {\AA} units these are: 0.30, 0.48, 0.78 (fivefold);
0.57, 0.92, 1.48 (twofold); and 0.20, 0.33, 0.53 
(threefold).

The densest fivefold layers of thickness less than
0.6\,{\AA} are pairs of planes 0.48\,{\AA} apart.
This agrees with experimental 
evidence~\cite{Gierer97,Z02}.
Fig.~\ref{fig:fig(3)}(I) graphs the atomic density,
$\rho(z_\perp)$, of $(q,b)$ pairs of planes as a
function of position along a fivefold $z_\perp$-axis
in coding space \es. It is close to its maximum along
a clear plateau of width about
$(2\tau^2/(\tau+2))$\,\ffo\,.
The interval of this plateau (a ``window'' $W$) codes
a Fibonacci sequence along a fivefold axis of \ep\ with
$S=(2\tau/(\tau+2))$\,\ffo${}=4.08$~{\AA} and 
$ L = \tau S = 6.60 $~{\AA}, in agreement 
with~\cite{G99}.
A slightly longer interval codes a decorated Fibonacci
sequence~\cite{Z02} that includes steps of height
$\tau^{-1}S=2.52$\,{\AA} too, detected in
Ref.~\cite{Shen99}.
Note that there is not such a clearly defined plateau
as in Fig.~\ref{fig:fig(3)}(I) when the windows in
Fig.~\ref{fig:fig(2)} are replaced by spherical
approximants.
As well as the $(q,b)$ layers there are also 
$(b,q)$ layers in \m\ whose density graph is the mirror 
image of Fig.~\ref{fig:fig(3)}(I) and codes another 
Fibonacci sequence of equally dense layers.
These are not seen as surface terminations,
the planes observed on the surface can be identified
as type $q$ by the presence of a local configuration
called a ``ring''~\cite{Z02}.
This can perhaps be understood from the densities of the 
parallel planes next above and below the layers: 
above a $(q,b)$ layer there is a low density
plane and  below a high density  plane.
For the mirror-image $(b,q)$ layer
the neighboring planes are mirrored too.
%
% Fig 3
%

The densest twofold layers are single planes.
In Fig.~\ref{fig:fig(3)}(II) we can identify a not
very sharply defined plateau of width about
\zfo/2 that defines the twofold terminations.
A window of this width on a twofold axis in \es\
encodes a Fibonacci sequence of terminations
along a twofold axis
of \ep\ with $S=(\tau^2/2)$\,\zfo${}=0.63$\,nm,
$L=\tau S=1.02$\,nm.
These gaps are close to the steps between terraces
(0.62\,nm and 0.95\,nm) in Fig.~\ref{fig:fig(1)}(II)(c).
Increasing the width of the window by a factor
$\tau^2$ would reduce the lengths of intervals in
the Fibonacci sequence by $\tau^{-2}$, making
them 0.24\,nm and 0.36\,nm corresponding to the
depths of the depressions in the terraces
in Fig.~\ref{fig:fig(1)}(II)(c).

The densest threefold layers are $(b,q,a)$ and $(a,q,b)$
triples of planes of with distances 0.20\,\AA\
between the $b$ and $q$ planes and 0.33\,\AA\
between the $q$ and $a$ planes.
The graph of the densities of the $(b,q,a)$ layers
as a function of position in coding space is shown
in Fig.~\ref{fig:fig(3)}(III) and has a wide
plateau-like area around its maximum.
This is in qualitative agreement with the
medium scale terraces that spread over an area of
about $1000 \times 1000$\,nm$^2$ on the threefold 
surface shown in Fig.\ref{fig:fig(1)}(III).
In the threefold case (unlike the fivefold case) there 
are single $b$ planes as dense as the densest threefold
terminations.

The second row of Table~\ref{tab:1} gives the
maximum densities of five, two and threefold 
terminations for the model \m, which are in the order 
fivefold, twofold, threefold.

We suggest that, with our modified definition of
``termination'', the Bravais rule that the densest
bulk terminations correspond to the most stable
surfaces may be widely applicable to quasicrystals.
Moreover the shape of the graph of the density of
layers as a function of position in coding space
determines the appearance of the corresponding surface:
if the maximum of the function has the form of a
flat plateau, as for the fivefold and threefold layers
in Figs.~\ref{fig:fig(3)}(I) and \ref{fig:fig(3)}(III), 
then the surfaces have a strong terrace-like character,
as in Figs.~\ref{fig:fig(1)}(I) and \ref{fig:fig(1)}(III).
The terraces correspond to bulk layers of almost
equal density and are equally probable as surfaces.
On the other hand, the sharper peak for the twofold
layers in Fig.~\ref{fig:fig(3)}(II) determines the more
fragmented appearance of the twofold surface with
small scale terraces as in Fig.~\ref{fig:fig(1)}(II)(a):
different terraces correspond to bulk layers
of different densities, and are not equally probable.

We have also observed that this  modified 
density rule is valid for \mbox{i-AlCuFe} and we expect 
that it should apply to all icosahedral quasicrystals. 
We have preliminary indications that this rule
can be extended to the so-called two-dimensional 
aperiodic materials (for example the family of
decagonal quasicrystals), and expect that such 
considerations may apply to the surfaces of quasicrystal 
approximant phases and other complex metallic alloys.

\end{document}